\documentclass[aps,prb,twocolumn,superscriptaddress,nofootinbib]{revtex4-1}
\pdfoutput=1
\usepackage{amssymb,graphicx,color}
\usepackage[intlimits]{amsmath}
\usepackage[english]{babel}
\usepackage[colorlinks]{hyperref}
\usepackage{comment}
\usepackage{tabularx}
\usepackage{array}
\usepackage[svgnames,table]{xcolor}
\usepackage{ragged2e}

\newcolumntype{A}{>{\centering\arraybackslash \columncolor{white!50!white}}m{2.1cm}}
\newcolumntype{B}{>{\centering\arraybackslash \columncolor{white}}m{7.9cm}}
\newcolumntype{C}{>{\centering\arraybackslash \columncolor{white!50}}m{7.9cm}}
\newcolumntype{D}{>{\centering\arraybackslash \columncolor{white!42}}m}
\newcolumntype{P}[1]{>{\centering\arraybackslash}p{#1}}
\def\beq{\begin{equation}}
\def\eeq{\end{equation}}
\def\bea{\begin{eqnarray}}
\def\eea{\end{eqnarray}}
\def\barr{\begin{array}}
\def\earr{\end{array}}

\begin{document}

\title{Stability of electric-field-driven MBL in an interacting long range hopping model}

\author{Devendra Singh Bhakuni}
\affiliation{Indian Institute of Science Education and Research Bhopal 462066 India}
\author{Auditya Sharma}
\affiliation{Indian Institute of Science Education and Research Bhopal 462066 India}

\begin{abstract}
 We study the fate of many-body localization (MBL) in the presence of
 long range hopping ($\sim 1/r^{\sigma}$) in a system subjected to an
 electric field (static and time-periodic) along with a slowly-varying aperiodic
 potential. We show that the MBL in the static electric-field model is
 robust against arbitrary long-range hopping in sharp contrast to
 other disordered models, where MBL is killed by sufficiently
 long-range hopping. Next, we show that the drive-induced phenomena
 associated with an ac square wave electric field are also robust
 against long-range hopping. Specifically, we obtain drive-induced
 MBL, where a high-frequency drive can convert the ergodic phase into
 the MBL phase. Remarkably, we find that a coherent destruction of MBL
 is also possible with the aid of a resonant drive.  Thus in both the
 static and time-periodic square wave electric field models, the
 qualitative properties of the system are independent of whether the
 hopping is short-ranged or long-ranged.
\end{abstract}

\maketitle 
\section{Introduction}
Many-body localization
(MBL)~\citep{basko2006metal,nandkishore2015many,PhysRevB.21.2366,PhysRevLett.95.206603},
in which localization is known to persist even in the presence of
many-body interactions, is a generalization of Anderson localization
~\citep{PhysRev.109.1492}. This is characterized by a number of
unusual features such as the area law scaling~\citep{bauer2013area}
for eigenstates far away from the ground state, logarithmic
time-evolution of entanglement
entropy~\citep{vznidarivc2008many,PhysRevLett.109.017202,PhysRevLett.110.260601}, and
violation of the eigenstate thermalization
hypothesis~\citep{PhysRevE.50.888,rigol2008thermalization}. Furthermore,
a system in the MBL phase is known to be robust against a high
frequency drive in contrast to generic many-body driven systems which
tend to a featureless `heat
death'~\citep{ponte2015periodically,lazarides2015fate,d2013many,abanin2015exponentially,ponte2015many, PhysRevB.93.235151,PhysRevB.94.020201,PhysRevB.94.094201}. Recent
advances on the experimental front have also helped to pile up
evidence for the MBL phase both in
static~\citep{schreiber2015observation,choi2016exploring,
  smith2016many,kondov2015disorder,bordia2017probing} as well as in
periodically driven systems~\citep{bordia2017periodically}.

Although the standard MBL phase is induced by random disorder,
systems with non-random potentials are also known to
exhibit the MBL phase~\citep{iyer2013many,lee2017many,modak2015many,
  nag2017many,li2015many}.  A recent proposal to realize MBL-like
features in a clean system subjected to a strong electric field and a
confining
potential~\citep{schulz2019stark,van2019bloch,bhakuni2020entanglement}
has created much excitement. Moreover, a time-periodic
electric field leads to many counter-intuitive
features~\citep{PhysRevB.102.024201,bairey2017driving} such as
drive-induced MBL and coherent destruction of Stark-MBL. In this
article, we explore the effect of static as well as a time-periodic
electric field on an interacting long-range hopping model, and discuss
the emergence of the MBL phase in this context. 

The study of the MBL phase in the presence of long-range
  forces has received considerable theoretical
  interest~\cite{PhysRevE.85.050102,nandkishore2017many,singh2017effect,roy2019self,PhysRevB.93.245427,PhysRevLett.113.243002,PhysRevB.97.214205,PhysRevA.93.022332,nag2019many,burin2006energy}.
  Long-range forces are ubiquitous in nature; recent technological
  advances, particularly in dipolar and trapped ionic
  systems~\cite{PhysRevLett.113.243002,smith2016many,korenblit2012quantum,roy2019quantum},
  have made the experimental realization of many of these systems, a
  realistic possibility. In systems with long-range hopping and
long-range interactions ($\sim 1/r^{\sigma}$), numerical studies have
shown that MBL cannot survive for
$\sigma<2$~\citep{PhysRevB.93.245427,PhysRevLett.113.243002,PhysRevB.97.214205,
  PhysRevA.93.022332}. Recently~\citep{nag2019many}, it was shown that
for a system with a quasi-periodic potential and short-range hopping,
the MBL phase is stable for arbitrary long-range interactions.
However, long-range hopping partially delocalizes the system for
$1<\sigma<2$ and completely delocalizes it for $\sigma<1$. This opens
up the question of how general these observations are, and whether a
stable MBL phase can be obtained for arbitrary long-range
hopping. Another question that springs up, in the backdrop of
drive-induced MBL~\citep{PhysRevB.102.024201,bairey2017driving} , is
the effect of a time-periodic electric field in the presence of such
long-range hopping. Here, we address these questions by studying an
interacting long-range hopping model subjected to a (static, and
time-periodic) electric field along with a slowly-varying aperiodic
potential and find that for suitable parameters, the MBL phase can
indeed be obtained for arbitrary long-range hopping.

Our main findings are as follows. For the un-driven model in the
short-range hopping case, we find that for small field strength and
slowly-varying aperiodic potential, the model shows an ergodic phase. In
contrast, for sufficiently large electric field and/or aperiodic
potential, the model shows an MBL phase where the nearest-neighbor
sites become entangled with each other. The nature of the MBL phase
itself seems to depend on whether the electric field or the
aperiodic potential is the dominant force. A study of the
dynamics shows a faster entanglement entropy growth in the MBL region
dominated by the aperiodic potential in comparison to regions
where MBL is induced by electric field. This suggests a strong
localization by the electric field. This localization remains robust
once the long-range hopping is turned on and a stable MBL phase is
obtained for a sufficiently large electric field, despite the presence
of an arbitrary long-range hopping. Strikingly, when the electric
field is converted into a time-dependent square pulse, we obtain both
drive-induced MBL and coherent destruction of MBL by a proper tuning
of the amplitude and frequency of the drive, even when the hopping is
long-ranged.

The organization of the paper is as follows. In Sec.~\ref{sec_2}, we
describe the model Hamiltonian along with the various static and
dynamical measures to distinguish the ergodic and the MBL phases. In
Sec.~\ref{sec_3}, we consider the case of a static electric field and
discuss the emergence of the MBL phase in the short-range and
long-range hopping limits. In Sec.~\ref{sec_4}, we discuss the effect
of a time-periodic electric field and the phenomenon of drive-induced
MBL in the long-range hopping case. Finally, we summarize our main
findings in Sec.~\ref{sec_5}.
\section{Model Hamiltonian and Quantifiers}\label{sec_2} 
We consider the Hamiltonian given by 
\begin{eqnarray}\label{eq1}
H=-\sum_{j>i}^{L-1}J_{ij}(c_{i}^{\dagger}c_{j}+h.c.)+\mathcal{F}(t)\sum_{i=1}^{L} i (n_{i}-\frac{1}{2}) \nonumber \\ -  \sum_{i=1}^{L} h_i (n_{i}-\frac{1}{2}) + V\sum_{i=1}^{L-1} (n_i-\frac{1}{2})(n_{i+1}-\frac{1}{2}),
\end{eqnarray}
where $\mathcal{F}(t)$ is a linear electric field, $V$ is the
nearest-neighbor interaction, and $J_{ij} = \frac{J}{|i-j|^\sigma}$ is
the long-range hopping. Since an additional confining
  potential is required for Stark-MBL, we perturb the electric field
  by an additional onsite potential: $h_i = h\cos(2\pi\beta i^n +
  \phi)$, where $\beta$ is some irrational number which we set to be
  the golden mean: $\beta = \frac{\sqrt{5}-1}{2}$ and $\phi\in(-\pi,
  \pi)$. We also set $n = 0.7$, thus making it a slowly-varying
  aperiodic potential, and adopt the natural units ($a=J=\hbar=e=1$).
To ensure the convergence of energy density in the thermodynamic
limit, we omit $\sigma<1$ in all the numerical
results~\citep{rodriguez2003anderson,balagurov2004phase,singh2017effect}.

The un-driven model is well-understood when it is short-ranged
$(\sigma\to\infty)$ and non-interacting $(V=0)$. This
  becomes a slowly varying aperiodic model for $\mathcal{F}(t)=F=0$,
  which, for $h<2J$~\cite{sarma1988mobility,sarma1990localization},
  shows a single particle mobility edge separating the localized and
  delocalized states. The states between energy $\pm|2J-h|$ are
  delocalized for $h<2J$, while all single-particle states are
  localized for $h>2J$. For $h=0$ and $F\ne 0$, we have the
well-known Wannier-Stark model in which all the single particle
eigenstates are localized and the eigen-spectrum forms an equi-spaced
ladder with the spacing proportional to the field
strength~\citep{krieger1986time,wannier1960wave}. As a consequnce of
the Wannier-Stark ladder, the dynamics shows Bloch
oscillations~\citep{hartmann2004dynamics,PhysRevB.98.045408,PhysRevB.99.155149}.  These
oscillations typically dephase in the presence of disorder and
non-linearity, however they are known to survive~\citep{de2005bloch}
in the presence of a slowly varying ($n < 1$) aperiodic potential.

In the presence of interactions $(V\ne 0)$ but short-range
hopping, either of electric field or a slowly-varying potential
separately  can yield both the ergodic and MBL phases for suitable tuning
of the field strength and the aperiodic
potential~\citep{schulz2019stark,van2019bloch, nag2017many}. In this
paper, we study the combined and relative effect of the electric field and the slowly-varying
aperiodic potential together in the presence of many-body interactions. We
further explore the scenario where the hopping is made long-ranged and
check the stability of the MBL phase. To do so we use the following quantifiers.

\subsection{Static measures: level spacing ratio and average nearest-neighbor concurrence} 
The ergodic and MBL phases can be characterized by
the average level spacing ratio, which is defined
as~\citep{oganesyan2007localization,atas2013distribution}
\begin{equation}
\langle r \rangle = \langle \frac{\text{min}(\delta_n,\delta_{n+1})}{\text{max}(\delta_n,\delta_{n+1})} \rangle,
\end{equation} 
where, $\delta_n = \epsilon_{n+1} - \epsilon_{n}$ is the gap between
subsequent energies. In the ergodic phase the average gap ratio
approaches the Gaussian orthogonal ensemble (GOE) value: $\langle
r\rangle = 0.53$ and satisfies Wigner-Dyson statistics whereas in the
MBL phase the gap ratio approaches the value : $\langle r\rangle =
0.386$ and satisfies Poisson statistics.

Similar to the average level spacing ratio, the average
nearest-neighbor concurrence, a measure of entanglement, also uniquely
characterizes the ergodic and the MBL phases~\citep{bera2016local} and
can be measured
experimentally~\citep{iemini2016signatures,PhysRevLett.115.035302,jurcevic2014quasiparticle}. From
the two-site reduced density matrix $\rho_{ij}$, the concurrence can
be calculated as
\begin{equation}
\mathcal{C}_{ij} =
\text{max}(0,\lambda^{(1)}-\lambda^{(2)}-\lambda^{(3)}-\lambda^{(4)}),
\end{equation}
where $\lambda^{(i)}$ are the eigenvalues of the matrix
$\sqrt{\rho_{ij}\tilde{\rho}_{ij}}$ arranged in the descending order
with $\tilde{\rho}_{ij}$ being the spin flip matrix and defined in
terms of Pauli spin matrix $\sigma^{y}$ as: $\tilde{\rho}_{ij} =
(\sigma^{y}\otimes \sigma^{y})\rho_{ij}^{*}(\sigma^{y}\otimes
\sigma^{y})$~\citep{wootters2001entanglement,chang2010dynamics}.
In systems with particle number conservation, the reduced density
matrix for the sites $i$ and $j$ reduces to a much simpler
form~\citep{nehra2018many,shu2005fermionic,chang2010dynamics}. Thus
the concurrence is given by
\begin{equation}
\mathcal{C}_{ij} =
2\text{max}(0,|z_{ij}|-\sqrt{u_{ij}v_{ij}}),
\end{equation}
where,
$u_{ij}=\langle(1-n_i)(1-n_j)\rangle, v_{ij}=\langle n_{i}n_{j}\rangle
\ \text{and}\ z_{ij}=\langle c_{j}^{\dagger}c_{i}\rangle$.

Here, we are interested in the entanglement between the
nearest-neighbor sites. We define the average nearest neighbor
concurrence as: $\mathcal{C} =\frac{1}{(L-1)}
\sum_{i=1}^{L-1}\overline{\mathcal{C}}_{i,i+1}$, where the over-bar
denotes the average over the central part of the eigen-spectrum. In
the ergodic phase, the volume law nature of eigenstates and the
principle of monogamy lead to a vanishing nearest-neighbor
concurrence while in the MBL phase, the area law nature of the
eigenstates together with the monogamy principle lead to a finite
concurrence between the nearest-neighbor
pairs~\citep{bera2016local,iemini2016signatures,da2017many}.

\subsection{Dynamical measures: Entanglement entropy, average concurrence and imbalance}
The distinction between the ergodic phase, Anderson localized phase,
and MBL phase can also be characterized by the post-quench dynamics of
various quantities. To study the dynamics of various
  observables, we employ the re-orthogonalized Lanczos
  algorithm~\citep{van2019bloch,luitz2017ergodic}.  Here, we focus on
the dynamics of the entanglement entropy, average nearest-neighbor
concurrence and the density imbalance starting with an initial state
where only the even sites are occupied.

The entanglement entropy is calculated from the density matrix by
identifying a subsystem and tracing out the degrees of freedom
complementary to it.  So, to study its time-evolution, we keep track
of the reduced density matrix as a function of time. In the present
study, we divide the system into two equal halves with each subsystem
having a length of $L/2$. For a given reduced density matrix
$\rho_{\text{red}}(t)$, the entanglement entropy can be calculated as:
$S(t) =
-\text{Tr}[\rho_{\text{red}}(t)\text{ln}\rho_{\text{red}}(t)]$. In the
ergodic phase, the growth of entanglement entropy is known to be
ballistic in time (barring some exceptions~\cite{luitz2016extended}
that show sub-linear growth) followed by a saturation to its thermal
value in the long time ~\citep{kim2013ballistic,de2006entanglement}
limit. In the MBL phase, the entanglement growth is unbounded and
logarithmic in time for an infinite system, while for finite systems,
it saturates to a sub-thermal value. In particular, it has been shown
that for the short-ranged XXZ model with weak interactions, the growth
of entanglement entropy in the MBL phase is given
by~\citep{PhysRevLett.110.260601} $S(t)\sim \xi\text{ln}(Vt/\hbar)$,
where $\xi$ is the single particle localization length.

Similar to the entanglement entropy, the dynamics of two-site
entanglement captured by the average concurrence helps identify the
different phases. Here, we consider the dynamics of the average
nearest-neighbor concurrence defined as: $\mathcal{C}_{NN}^{T}(t) =\frac{1}{(L-1)}
\sum_{i=1}^{L-1}\mathcal{C}_{i,i+1}(t)$.  The average concurrence is
known to decay rapidly in the ergodic phase as the entanglement
spreads very fast due to much faster propagation of the quasi-particle
excitations. This means that the nearest-neighbor concurrence will
vanish rapidly as suggested by the monogamy of entanglement. In the
Anderson localized phase, the average concurrence increases for a
short time and then saturates, while in the MBL phase, after increase
for a short time it shows a power-law decay~\citep{iemini2016signatures}. 

Finally, we consider the
dynamics of the density imbalance with which signatures of the MBL
phase has been obtained
experimentally~\citep{schreiber2015observation,bordia2017probing}. The density
imbalance is defined as
\begin{equation}
\mathcal{I}(t) = \frac{\mathcal{N}_{e}(t) -
	\mathcal{N}_{o}(t)}{\mathcal{N}_{e}(t) + \mathcal{N}_{o}(t)},
\end{equation}
where $\mathcal{N}_{e}(t)$ and $\mathcal{N}_{o}(t)$ are respectively the
average number of particles at the even and the odd sites at any
instant of time $t$. In the ergodic phase, the imbalance quickly
saturates to zero, thus suggesting that it is unable to retain the
memory of the initial state. In the MBL phase on the other hand, the
imbalance saturates to a non-zero value and retains the memory of the
initial state.
\section{Static Electric field}\label{sec_3} 
\subsection{MBL in short-range limit $(\sigma\to\infty)$}
\begin{figure}[t]
	\includegraphics[scale=1.17]{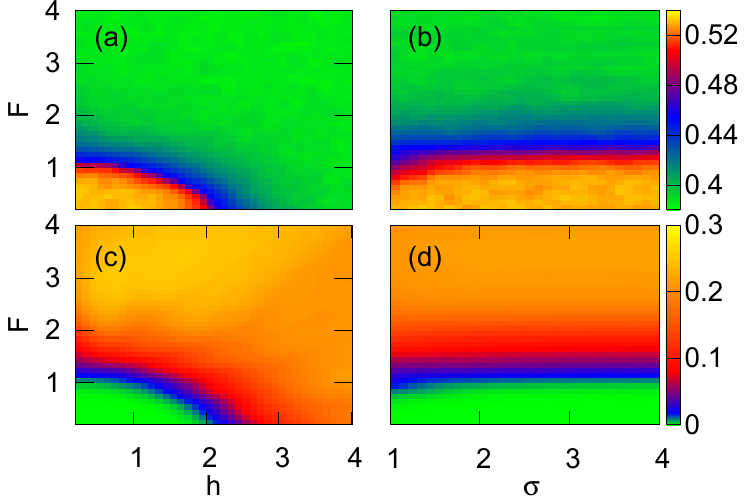}
	\caption{Left panel: (a,c) Surface plot of the average level
          spacing ratio and average nearest-neighbor concurrence as a
          function of both field strength and the aperiodic
          potential for the short-ranged model. Right panel: (b,d)
          Both level-spacing ratio and average concurrence as a
          function of field strength and long-range parameter $\sigma$
          for a fixed aperiodic potential $h=0.2$. The average
          nearest-neighbor concurrence is zero in the ergodic phase,
          while a non-zero average concurrence is obtained for the MBL
          phase. The other parameters are: $L=16, V=1.0, n=0.7$. For
          calculating the average nearest-neighbor concurrence, the
          average is carried out over $3000$ eigenstates from the
          central part of the spectrum. }
	\label{concSR}
\end{figure}

In the short range limit $(\sigma\to\infty)$, we have an interacting
nearest-neighbor hopping model subjected to a static electric field
and a slowly varying aperiodic potential. We first consider the
average level spacing ratio to characterize the different phases. The
value of $\langle r\rangle$ is plotted in Fig.~\ref{concSR}(a) as a
function of both field strength and the aperiodic potential. An
MBL phase is obtained for sufficiently large electric field strength
and aperiodic potential.  A similar trend can be seen from the
average nearest-neighbor concurrence (Fig.~\ref{concSR}(c)) which is
obtained by averaging over $3000$ eigenstates (for a system of size
$L=16$ at half-filling) lying in the central part of the spectrum. The
concurrence vanishes in the ergodic phase where the volume law nature
of the states implies a vanishingly small entanglement between the
neighboring sites. On the other hand, in the MBL phase, the
nearest-neighbor pairs acquire a non-zero entanglement.
\begin{figure*}[t]
	\includegraphics[scale=0.775]{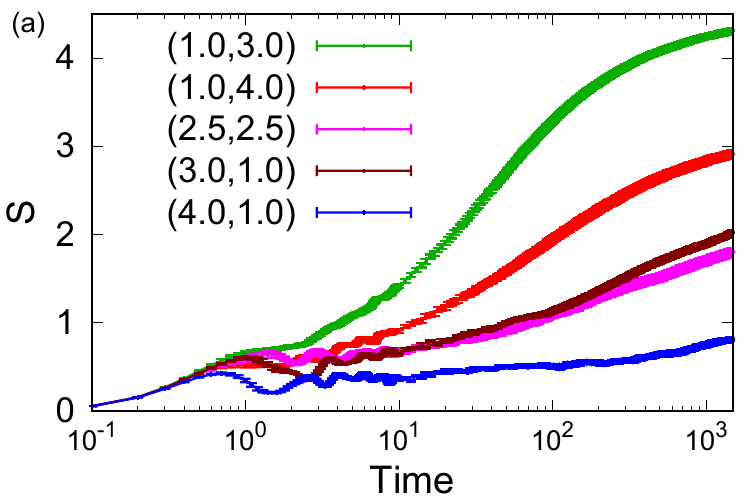}
	\includegraphics[scale=0.775]{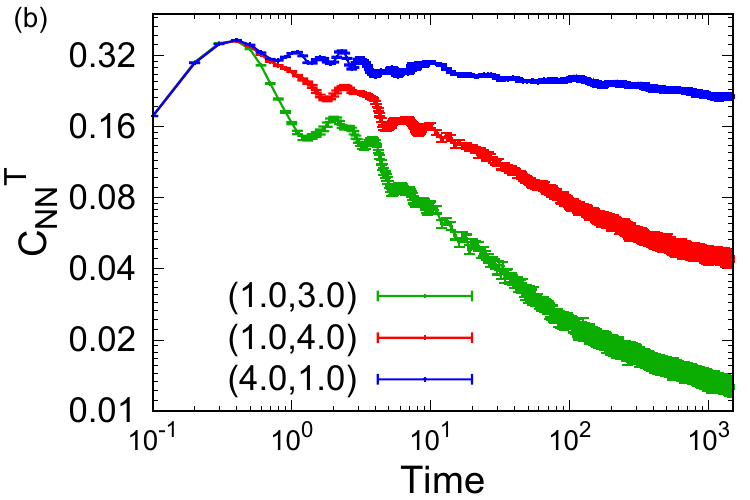}
	\includegraphics[scale=0.775]{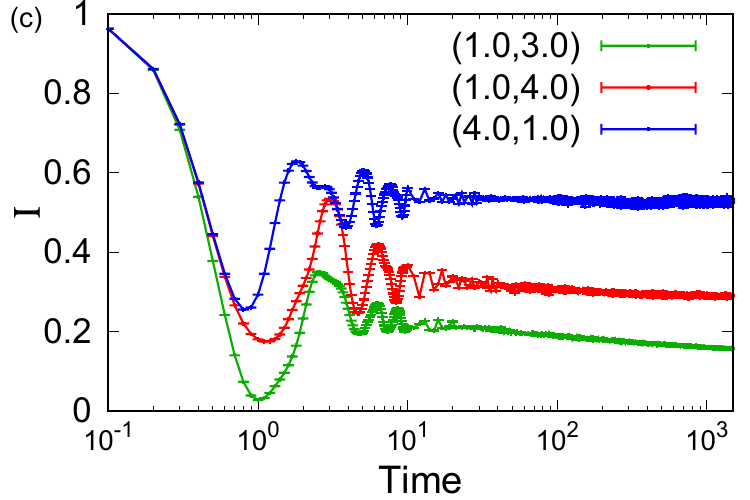}
	\caption{Short-ranged model: The dynamics of $(a)$
		entanglement entropy, $(b)$ average nearest-neighbor
		concurrence, and $(c)$ imbalance for various values of
		electric field strength and aperiodic potential. The
		legend shows the value of electric field strength and the
		strength of aperiodic potential $(F,h)$. Although for
		all parameters the system is in the MBL phase, the growth of
		entanglement entropy (and decay of average nearest-neighbor
		concurrence) is quite different for different points in the
		MBL phase. The other parameters are: $L=20, V=1.0,
		n=0.7$. The data are averaged over $100$ configurations of
		the phase $\phi$. }
	\label{entimbSR}
\end{figure*}
\begin{figure*}[t]
	\includegraphics[scale=0.775]{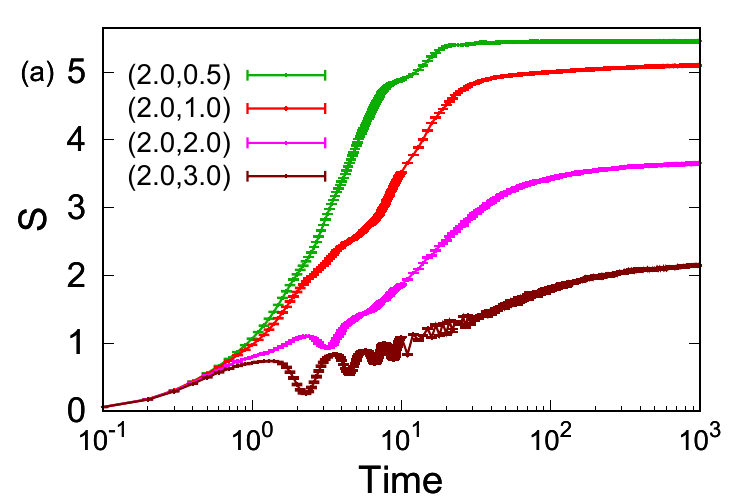}
	\includegraphics[scale=0.775]{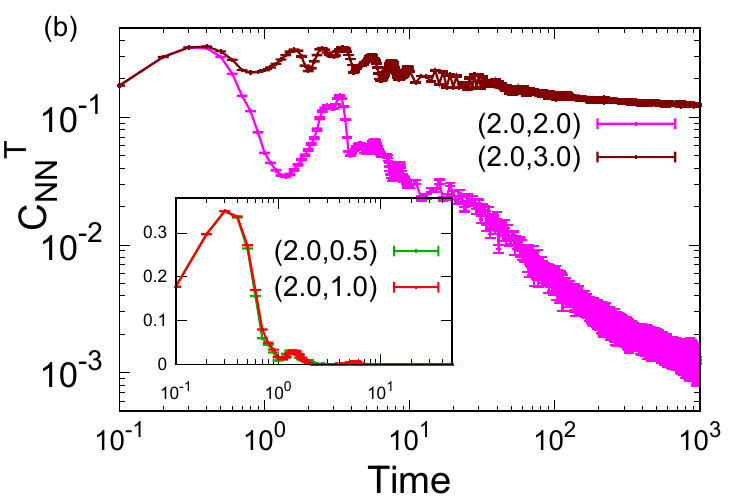}
	\includegraphics[scale=0.775]{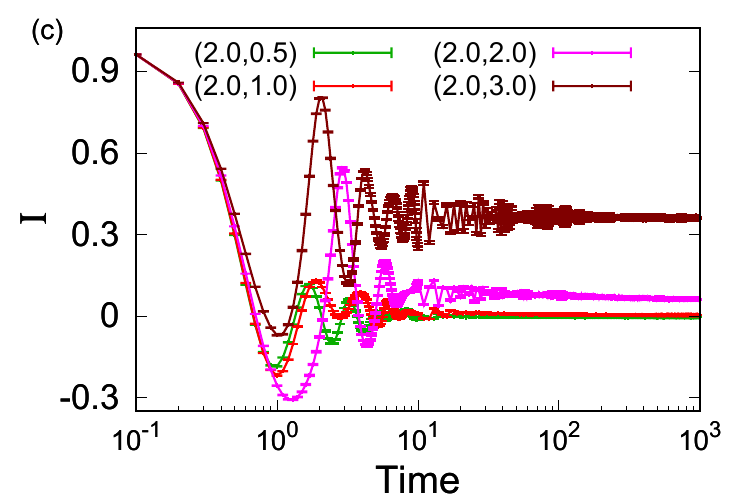}
	\caption{Long-ranged model: The dynamics of $(a)$ entanglement
		entropy, $(b)$ average nearest-neighbor concurrence, and
		$(c)$ imbalance for various values of electric field
		strength and long-range parameter $\sigma$. The legend shows
		the value of the long range parameter and electric field
		strength $(\sigma,F)$.  The inset in $(b)$ shows the
		dynamics of average nearest-neighbor concurrence for the
		parameters in the ergodic phase, where it rapidly decays to
		zero.  The strength of the aperiodic potential is kept
		fixed at $h=0.2$. The other parameters are: $L=18, V=1.0,
		n=0.7$. The data are averaged over $100$ configurations of
		the phase $\phi$.}
	\label{entimbLR}
\end{figure*}

We then move to the dynamic characterization of the ergodic and MBL
phases. The study of dynamics reveals a quantitative difference
between the MBL phase obtained from the electric field (plus small
aperiodic potential) and the MBL phase obtained from the
aperiodic potential (with small field
strength). Fig.~\ref{entimbSR}(a) shows the dynamics of entanglement
entropy from the aforementioned initial state consisting of alternate
filled and empty sites. The plot shows the growth of entanglement
entropy (averaged over $100$ different configurations of $\phi$) for
the various points ($F,h$) in the MBL phase. It can be seen that for
the chosen parameters, the initial (upto $t\approx 1$) growth of the
entanglement is the same for all the parameters and after this the
entanglement entropy grows logarithmically in time. The MBL phase
obtained from the electric field is characterized by a slower
logarithmic growth as compared to the MBL phase obtained from the
aperiodic potential signifying a stronger localization by the
electric field as compared to that by the aperiodic potential.

The dynamics of the average nearest-neighbor concurrence is also
plotted in Fig.~\ref{entimbSR}(b). Similar to the entanglement
entropy, the initial growth (upto $t\approx 1$) is the same for all
the parameters, while it starts to decay as a power-law for longer
times. Again, the decay is fast for the points where the MBL is
dominated by the aperiodic potential, while it has a slow decay
for the MBL dominated by the electric field. Similarly a study of the
dynamics of imbalance (Fig.~\ref{entimbSR}(c)) shows that it saturates
to a higher value for MBL induced by the electric field as opposed to
the aperiodic driven MBL. The dynamical measures provide a
hint that MBL induced by electric field can be more robust against any
perturbations. We explicitly check this in the next subsection by
turning on long-range hopping in the system.

\subsection{MBL for long-range hopping} 
Next, we study the stability of MBL in the presence of arbitrary
long-range hopping. Fig.~\ref{concSR}(b,d) show the surface plots of
the average level spacing ratio and the average nearest-neighbor
concurrence respectively as a function of both field strength and the
long-range parameter $\sigma$ for a fixed aperiodic potential
($h=0.2$). It can be seen from Fig.~\ref{concSR}(b) that for small
values of the electric field and all values of the parameter $\sigma$,
the level-spacing ratio satisfies Wigner-Dyson statistics and thus
signifies an ergodic phase, whereas for large field strength it
satisfies Poisson statistics and an MBL phase is indicated. Similarly,
the average nearest-neighbor concurrence (Fig.~\ref{concSR}(d)) is
zero for small values of the electric field and for arbitrary
long-range hopping. In contrast, for large electric fields, the
average concurrence shows a non-zero value for any arbitrary
long-range hopping. It is worth noting that the magnitude of the
electric field where the transition between the MBL and ergodic phases
happens is almost the same as that of the short-range
model (Fig.~\ref{concSR}(a,c)). This leads to the conclusion that the
MBL induced by the electric field (in the short-range limit) is robust
in the presence of arbitrary long-range hopping.
\begin{figure*}
	\includegraphics[scale=0.33]{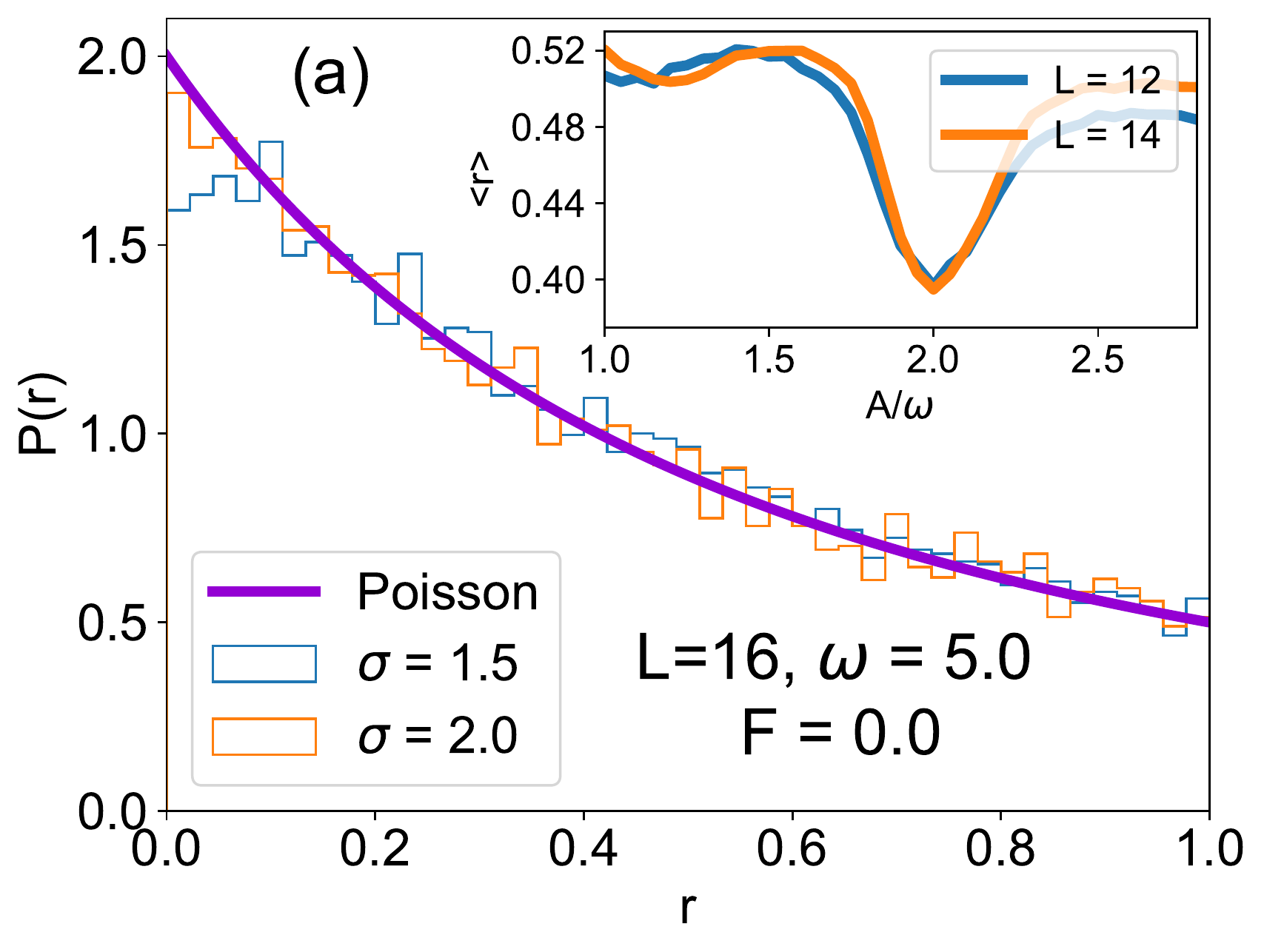}
	\includegraphics[scale=0.33]{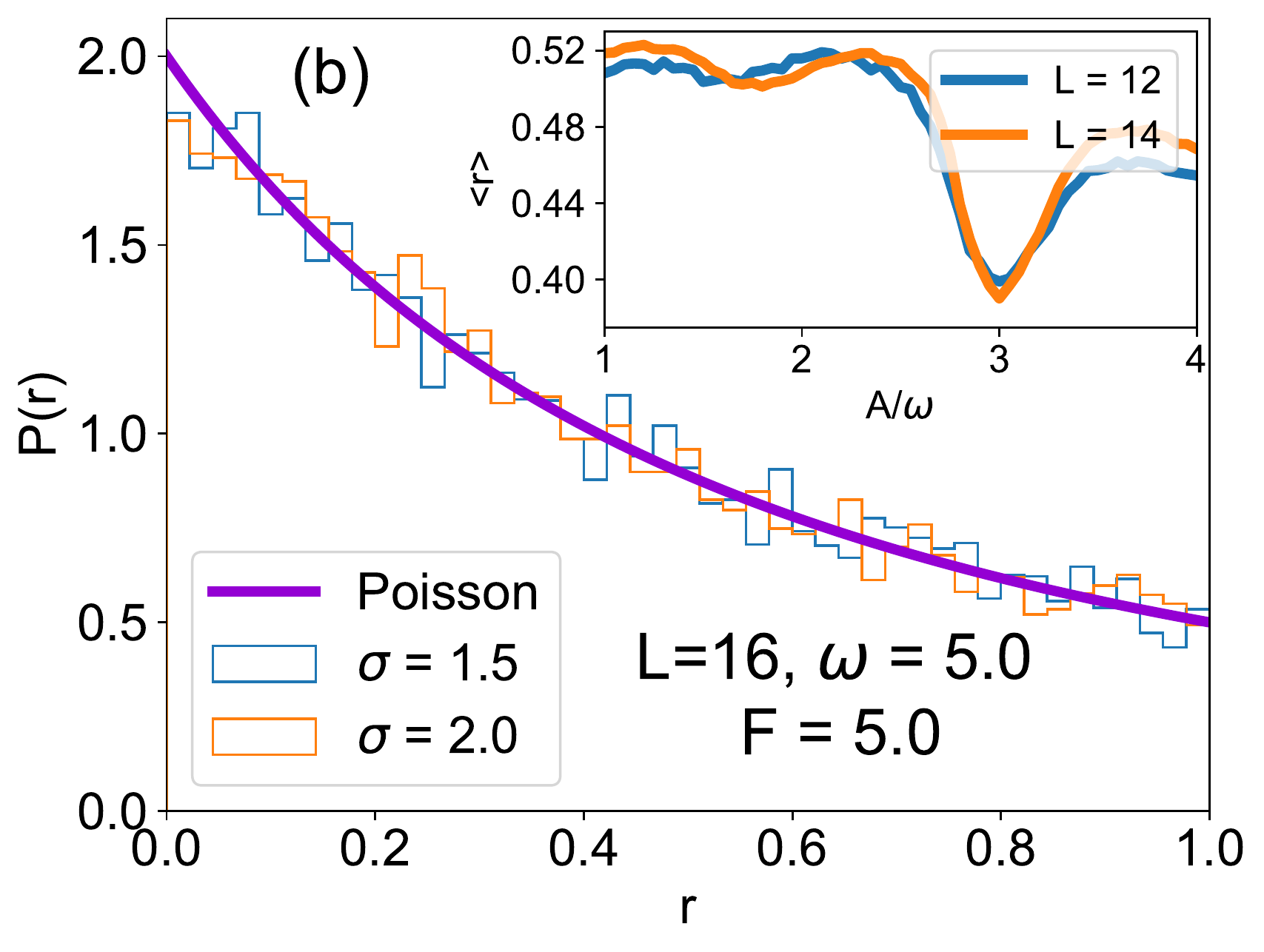}
	\includegraphics[scale=0.33]{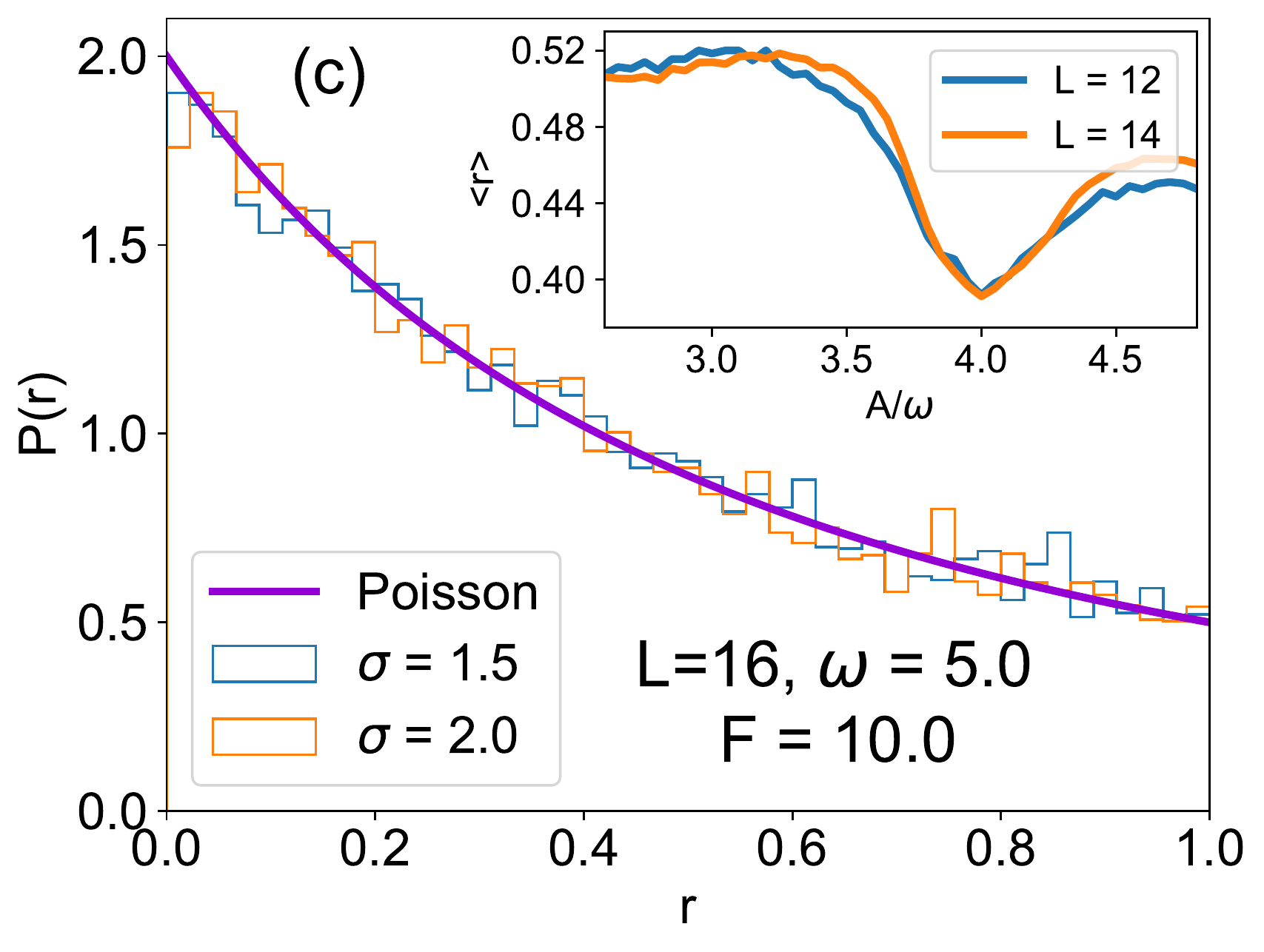}
	\caption{(a-c) Probability distribution of the quasi-energy
          gap ratio parameter for a high frequency drive
          ($\omega=5.0$) at the dynamical localization point. The
          cases:$(a)$ without any static field, $(b)$ resonantly tuned
          drive ($F=n\omega$) with $n$ odd and $(c)$ resonantly tuned
          drive with $n$ even are shown. The insets in (a-c) show the
          average level-spacing ratio (averaged over $100$
          configurations of $\phi$) for $\sigma = 1.5$, as a function
          of the ratio $A/\omega$ for different system sizes. In all
          the cases, the MBL phase is found by tuning the ratio at the
          dynamical localization point, whereas away from the
          dynamical localization point an ergodic phase is observed.}
	\label{ravqeig}
\end{figure*} 

We then look at dynamical quantities to characterize the MBL and the
ergodic phases. Fig~\ref{entimbLR}(a) shows a plot of the dynamics of
the entanglement entropy for $\sigma =2$ and various values of the
field strength. After an initial growth, the entanglement entropy
saturates to a finite value for small field strengths. The saturation
value of the entropy for $F=0.5$ and $L=18$, is $S(t=1000)=5.45$,
which is close to the thermal value predicted by
Page~\citep{page1993average} : $S(t\to\infty) = \frac{L}{2}\text{ln}2
- 0.5$ when the subsystem consists of half of the sites $(L/2)$. For
larger field strength $F=3.0$, the entropy displays a slow
growth followed by saturation, thus signifying a stable MBL phase
for larger field strengths. The average nearest-neighbor concurrence
is also plotted in Fig.~\ref{entimbLR} (b). Again for small field
strengths, the average nearest-neighbor concurrence is vanishingly
small (inset of Fig.~\ref{entimbLR}(b)), while for larger field
strengths it decays as a power-law which supports the argument that
for large $F$, a stable MBL phase is observed. Furthermore, the
dynamics of the density imbalance also suggests a stable MBL phase for
larger field strengths where it saturates to a finite non-zero
value. The saturation value decreases on decreasing the field strength
where we would expect localization to be less strong.

\section{Time-periodic electric field}\label{sec_4}
In this section, we consider a more general electric field that has
both dc as well as ac components and restrict to the long-range
hopping case. We build on a previous work
~\citep{PhysRevB.102.024201}, where we considered the short-range
version of a variant Hamiltonian. The electric field can be written
as: $\mathcal{F}(t) = F +A\;\text{sgn}(\sin(\omega t))$, where $A$ and
$\omega$ respectively are the amplitude and frequency of the ac field,
while $F$ is the static dc field. A pure ac square wave drive (with $F
= 0$ above) yields dynamical localization even in the presence of
arbitrary long-range hopping in contrast to sinusoidal
driving~\citep{Dunlap1986,dignam2002conditions}, where long-range
hopping destroys dynamical localization. We will first consider the
non-interacting case, where we obtain the condition for dynamical
localization for a combined square wave ac and dc field.

In the absence of both the electric field and the aperiodic
potential, the dispersion of the Hamiltonian is given by:
\begin{equation}
E(k)=-2\sum_{m>0}J_m\cos mk,
\end{equation} 
where $J_m = 1/m^\sigma$. When the field is turned on, the
quasi-momentum can be written as: $q_{k}(t)=k+Ft+\int_{0}^{t} d\tau
F_{ac}(\tau)$, where $F_{ac}$ is the ac part of the field. Due to the
dc part, the quasi-momentum is a periodic function only if the
resonance condition: $F=n\omega$, holds. Following the same
prescription as in the short-ranged model~\citep{PhysRevB.102.024201},
the quasi-energy is given as
\begin{equation}\label{8}
\epsilon(k) = -2\sum_{m>0}J_{m}^{\text{eff}}\cos(mk+\frac{mn\pi}{2}),
\end{equation}
where 
\begin{equation}
J_{m}^\text{eff} = J_{m}\left\lbrace\frac{\sin(\frac{mn\pi}{2}+\frac{mK\pi}{2})}{m(K\pi+n\pi)}+(-1)^{n}\frac{\sin(\frac{mK\pi}{2}-\frac{mn\pi}{2})}{m(K\pi-n\pi)}\right\rbrace,
\end{equation}
and $K =A/\omega$. 

For even and odd $n$ respectively,
we get $\epsilon(k) = -2 \sum_{m>0} J_{m}^{e} \cos(mk)$ and $\epsilon(k) = -2 \sum_{m>0} J_{m}^{o} \sin(mk)$, where
\begin{equation}\label{evenodd}
J_{m}^{e}=\frac{2J_{m}K\sin(\frac{mK\pi}{2})}{m(K^{2}-n^2)\pi};\;\;J_{m}^{o}=\frac{2J_{m}K\cos(\frac{mK\pi}{2})}{m(K^{2}-n^2)\pi}.
\end{equation}
Thus the effect of drive in this case is again the renormalization of
the hopping parameter. For the even and odd cases respectively, we
obtain the condition for dynamical localization by looking at the
band collapse points which occur at $K = K_{c}=2\nu$ and $K =
K_{c}=2\nu +1$, $\nu$ being any integer and $K_{c}\neq n$. At these
points, the band collapse forces an initially localized wave-packet to
return to its initial position. However, for other values of $K$, a
delocalization effect can be seen due to the band formation. This also
provides the mechanism of coherent destruction of Wannier-Stark
localization but for an arbitrary long-range hopping.

In the presence of many-body interactions, this dynamical localization
is destroyed. However, if an additional small aperiodic potential
is present, the MBL phase is obtained close to the dynamical
localization point. Fig.~\ref{ravqeig} shows the probability
distribution of the gap-ratio parameter for different values of the
static field strength and long-range parameter. In all the cases,
the probability distribution matches with that of a Poisson
distribution, which suggests the MBL phase at these values. The insets
in Fig.~\ref{ravqeig} show the average gap-ratio as a function of
$A/\omega$. It can be seen that only around the dynamical localization
point, an MBL phase is obtained, while for other values of $A/\omega$,
an ergodic phase is observed. Similar to the short-range
case~\citep{PhysRevB.102.024201}, the obtained results can be
interpreted as follows: for zero dc field and a small aperiodic
potential ($h=0.2$), the un-driven model is in the ergodic phase
(Fig.~\ref{concSR}). The application of a high-frequency drive in this
case leads to the MBL phase around the dynamical localization point
and an ergodic phase away from the dynamical localization point. This
is again a case of drive-induced many-body localization, but for an
arbitrary long-range hopping.

Similarly, for a non-zero dc electric field ($F=5.0,10.0$), where the
un-driven system is in the MBL phase, the application of a resonant
drive can destroy it provided that the ratio $A/\omega$ is tuned away
from the dynamical localization point (Fig.~\ref{ravqeig}(b,c)). At
the dynamical localization point, the MBL phase is obtained.
  
\section{Summary and conclusions}\label{sec_5}
To summarize, we study an interacting long-range hopping model
subjected to a general electric field (both ac and dc) and a slowly
varying aperiodic potential.  We start by studying the short-range
hopping case of the undriven model, and find that for small electric
field strength and aperiodic potential, the model shows the ergodic
phase while for sufficiently large electric field and the aperiodic
potential, the model shows the MBL phase. However, the MBL phase
obtained by cranking up the electric field is quite different from the
MBL phase obtained from large aperiodic potential. While the MBL phase
obtained with a large electric field strength and small aperiodic
potential shows a slow growth of entanglement entropy, the MBL phase
arising from a large aperiodic potential shows a fast growth of
entanglement entropy. The distinction becomes even clearer once
the hopping is made long-ranged. We find a stable MBL phase for a
sufficiently large electric field. This is to be contrasted with the
result~\citep{nag2019many} that with a dominant quasi-periodic
potential, a transition from ergodic to MBL phase is obtained on
varying the long-range parameter.

By turning on drive in the long-range hopping model, we obtain
generalized conditions for dynamical localization for a combined dc
and ac field in the non-interacting limit. In the presence of
interactions and small aperiodic potential, we find a
high-frequency drive-induced MBL phase for an arbitrary long-range
hopping by tuning the system at the dynamical localization point. On
the other hand, the drive is also found to take the MBL phase of the
un-driven model to the ergodic phase. This shows that the coherent
destruction of Stark-MBL is possible, even in the presence of an
arbitrary long-range hopping. Thus for both the static and
time-periodic square wave electric field, the qualitative behavior of
the system are found to be independent of whether the hopping is
short-ranged or long-ranged.  This fact may help towards the
experimental detection of the MBL phase in long-ranged systems.

\section*{Acknowledgments}
We thank Fabien Alet for his insightful comments on the manuscript.
We are grateful to the High Performance Computing(HPC) facility at
IISER Bhopal, where large-scale calculations in this project were
run. A.S ackowledges financial support from SERB via the grant (File
Number: CRG/2019/003447), and from DST via the DST-INSPIRE Faculty
Award [DST/INSPIRE/04/2014/002461]. D.S.B acknowledges PhD fellowship
support from UGC India.

\bibliography{ref}
\onecolumngrid
\appendix
\section{Finite-size flow}
\begin{figure}[!h]
	\includegraphics[scale=0.775]{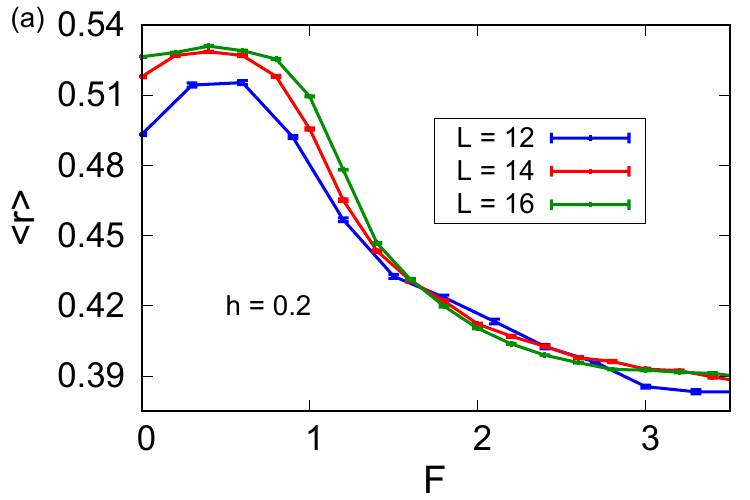}
	\includegraphics[scale=0.775]{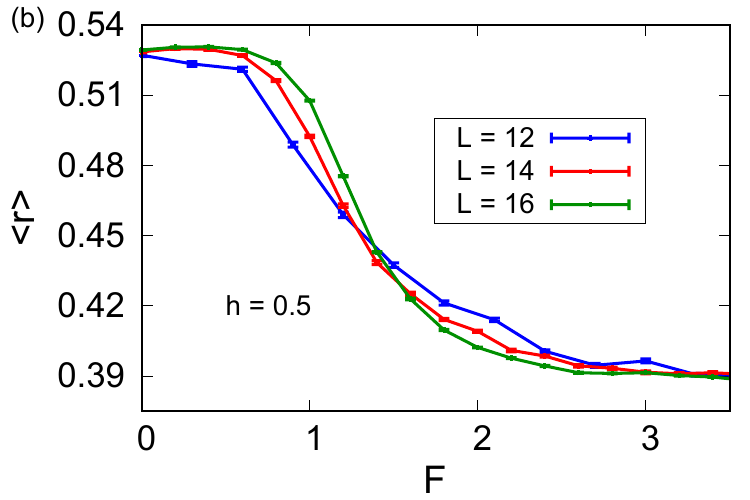}
	\includegraphics[scale=0.775]{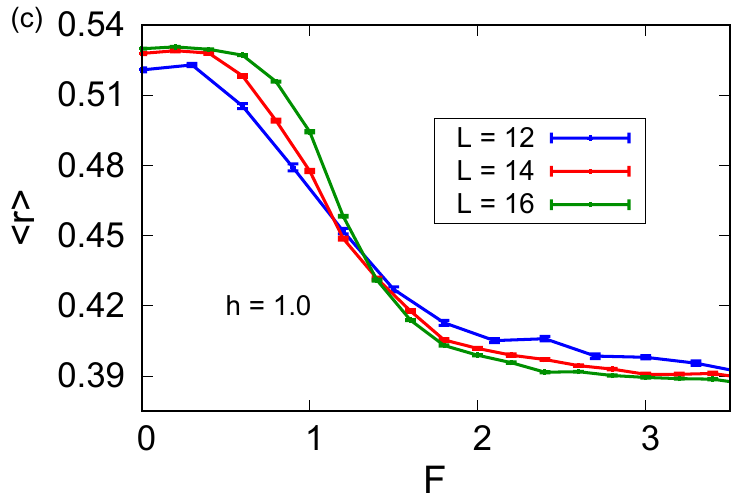}
	\caption{Finite-size flow: (a-c) Average level spacing ratio
          as a function of the field strength in the short-range limit for $h=0.2,0.5$ and $1.0$ respectively.
          It can be seen that in all the cases,
          for small values of the electric field, an ergodic phase is
          obtained, while for large field strengths an MBL phase is
          obtained. The data
          are averaged over $150$ configurations of the phase $\phi$.}
	\label{ravFS}
\end{figure}

In this section, we carry out a finite-size flow analysis of
the obtained phases in the main text. We first consider the
short-range case. Fig.~\ref{ravFS}(a) shows the level-spacing ratio as
a function of the field strength for a fixed strength of the aperiodic
potential $h=0.2$. The finite-size supports the
conclusion that an MBL phase is obtained for sufficiently large field
strength, while an ergodic phase is obtained for small field strength. Although, the data for different system sizes seem to cross at various points, we suspect this due to a smaller value of the aperiodic potential ($h=0.2$) which is unable to entirely break the many-body degeneracies~\cite{schulz2019stark} of the problem at very large field strength. To check this explicitly, we present the data for different values of the aperiodic potential ($h=0.5,1.0$, Fig.~\ref{ravFS}(b,c)) which shows only a single crossing. 

\begin{figure}[!h]
	\includegraphics[scale=0.775]{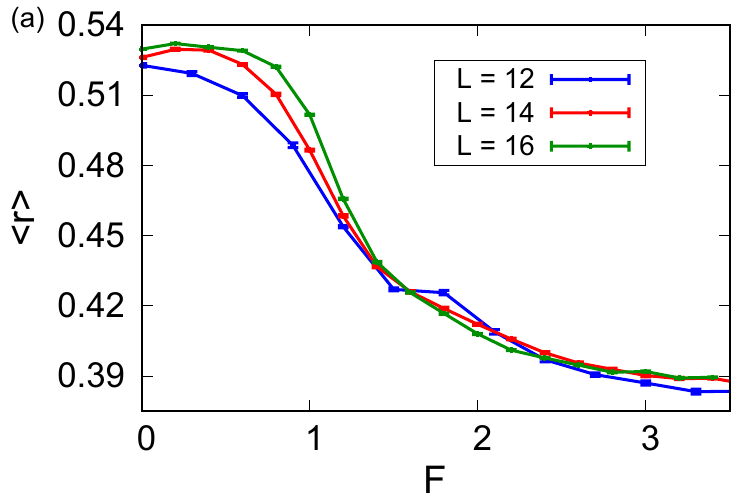}
	\includegraphics[scale=0.775]{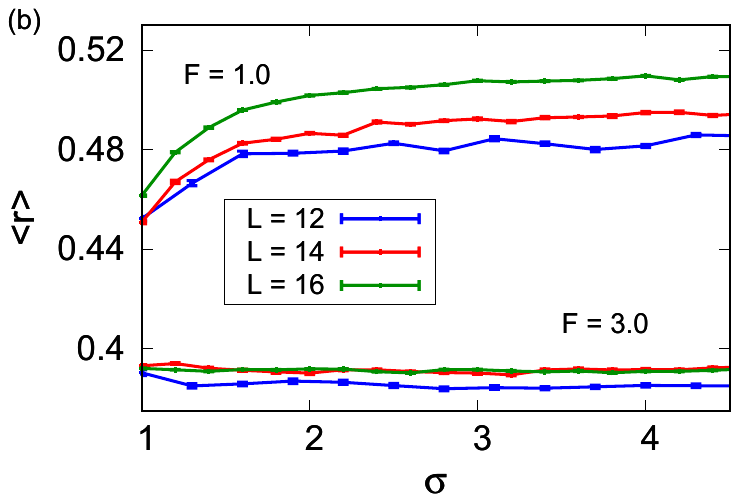}
	\caption{Finite-size flow: (a) Average level spacing ratio
		as a function of the field strength in the long-range hopping
		case ($\sigma=2.0$). An MBL phase is
		obtained only for sufficiently large field strengths. (b) The average level-spacing ratio as a function
		of the long-range parameter $\sigma$ for $F=1.0, 3.0$ and
		$h=0.2$. An MBL phase is obtained for arbitrary long-range
		hopping only at sufficiently large field strengths. The data
		are averaged over $150$ configurations of the phase $\phi$.}
	\label{ravFSL}
\end{figure}

We then move to the long-range hopping case with $\sigma = 2.0$, and
study the variation of the average level-spacing ratio as a function
of the field strength with a fixed aperiodic potential
($h=0.2$). Again, finite-size flow suggests an ergodic phase for
small field strength and an MBL phase for large field strength
(Fig.~\ref{ravFSL}(a)). Finally, we fix the field strength ($F = 1.0,
3.0$) and vary the long-range parameter $\sigma$
(Fig.~\ref{ravFSL}(b)). The obtained results lead to the conclusion
that the Stark-MBL phase is stable for arbitrary long-range hopping.

\section{AAH Limit ($n=1$)}  
\begin{figure}[!h]
	\includegraphics[scale=2.2]{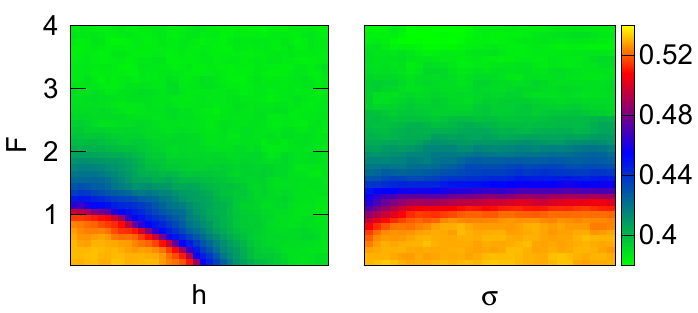}
	\caption{Left: Surface plot of the average level spacing ratio
          as a function of both field strength and the quasi-periodic
          potential for the short-ranged model in the AAH limit
          ($n=1.0$). Right: Surface plot of the average level-spacing
          ratio as a function of field strength and the long-range
          parameter for a fixed $h=0.2$. The other parameters are:
          $L=16, V=1.0, n=1.0$. }
	\label{ravqpSR}
\end{figure}

In this section, we consider the limit ($n=1$) where in the absence of
the field strength, the non-interacting short-range model becomes the
well known Aubry-Andre Harper (AAH) model. Fig.~\ref{ravqpSR} shows
the surface plots of the average level spacing ratio for $n=1$,
similar to the plots shown in Fig.~\ref{concSR}. It can be seen that
the qualitative behavior is similar to Fig.~\ref{concSR}. This is
expected as the electric field dominates over the quasi-periodic
potential for small quasi-periodic/aperiodic potential strengths while
the qualitative behavior is same for both the quasi-periodic and the
aperiodic potential for very high values of $h$.
\end{document}